# Broadband Low-loss Unidirectional Reflection On-chip with Asymmetric Dielectric Metasurface

Heijun Jeong[1*†], Zeki Hayran[2†], Yuan Liu[3], Yahui Xiao[1], Hwaseob Lee[1], Zi Wang[1], Jonathan Klamkin[3], Francesco Monticone[2*] and Tingyi Gu[1*]

[1] Electrical and Computer Engineering, University of Delaware, Newark, DE 19716

[2] Electrical and Computer Engineering, Cornell University, Ithaca, NY 14853

[3] Department of Electrical and Computer Engineering, University of California, Santa Barbara, CA 93106, USA

*Email: heijun@udel.edu, francesco.monticone@cornell.edu, tingyigu@udel.edu

†Authors contributed equally

**ABSTRACT.** Metasurface has emerged as a powerful platform for controlling light at subwavelength thickness, enabling new functionalities for imaging, polarization manipulation, and angular momentum conversion within a flat surface. We explored an integrated asymmetric metasurface simultaneously achieving broadband, low loss forward power transmission, and significant back reflection suppression in multi-mode waveguides. The tapering along the direction of light propagation leads to low loss and space-efficient mode conversion. Enhanced by a double-flipped structure, a thin (2.5 µm) metasurface can simultaneously achieve high conversion efficiency (>80%), and back-reflection efficiency of 90% over a 200 nm wavelength range. Such single-side reflectors can be one of the enabling components for gain-integrated adaptive optics on a chip.

# INTRODUCTION

Introducing the phase non-uniformity and discontinuity in single and multi-layer metasurface allows the exploration of many intriguing physical phenomena[1–5]. Parallel to the time-domain modulation of non-Hermitian photonic systems[6], parity-symmetry can be engineered through the subwavelength taper along the direction of light propagation[2]. Simple gliding and offsets between double-layer metallic metasurface enhance sensing through plasmonic exceptional points[7,8]., or BIC in dielectric meta-layers[9–11]. Relative rotation at a magic angle engineers the level of light localization and the formation of photonic Moiré lattices[12,13], and hyperbolic surface[14]. Stacking muti-layer meta-structure with a specific time reversal symmetry design variation along the direction of out-of-plane incident wave constructs photonic topological isolator metacrystal[15], Hofstadter butterfly, and topological edge states[1,16,17]. Asymmetric meta-atoms in free-space optics, typically with broken in-plane symmetry, can support high $Q$ resonance through symmetry-protected bound states in the continuum[18–22], provide arbitrary polarization[23] and phase control[24], or angular momentum conversion[25]. Note that such geometric symmetry cannot replace the magnetic material-based BIC structure for stable generation of C points with broken time-reversal symmetry[26,27]. Engineering the non-Hermiticity of metasurface and their complex eigenvalues through phase gradience allows the directional coupling of the input light to the radiation channels[2,5,28,29].

Phase-varying asymmetric metasurface allows the asymmetric phase shift for the forward and back-excited modes[2]. The broken spatial parity symmetry leads to distinguished transmission between normally incident forward and backward waveguide channels near the phase singularities in a complex plane. Currently, related structures have been explored in free space optics systems, where the geometric offset along the light propagation direction needs careful inter-layer alignment. In guided wave circuits, more sophisticated and finely adjusted geometric variations along the wave's propagation direction can be introduced through easy and one-step lithography. Tilted meta structures led to low loss and low crosstalk mode conversion in dielectric waveguides[30–32] and surface plasmon polaritons[33,34]. Inverse-designed integrated metalens with nanoscale topology optimization trimmed the wavefronts for compact mode size conversion[35,36]. Here we explored the design concept in multi-mode waveguides, with applications of simplifying the circuit design complexities used for ion trapping[37,38], recurrent photonic neuron networks[39,40], and high-power RF-photonic systems[41]. Here we show that the unit-cell level longitudinal asymmetry is critical for broadband and low-loss transmission while keeping nearly unit reflection for the back-reflected light. Also, we provided examples of how the compact passive back reflector can be applied to suppress the broadband back reflections from the random backscattering from material nonuniformity or surface roughness.

**RESULTS**

Here we focus on the study of metasurface asymmetry along the direction of propagation, with the taper length kept within the dimension comparable to the effective wavelength in the media. The forward and backward excited waves experience different phase shifts. Previous works have shown that arrays of rectangular meta-units (silicon oxide void) in silicon photonics can deterministically control wavefronts with low insertion loss[35,36]. Phase-varying asymmetric metasurface allows the asymmetric phase shift for the forward and back-excited modes. Here we show asymmetric power transmission and momentum transfer can be achieved through one-side tapering of the slot width (Fig. 1a). The transmission contrast increases with geometric contrast ($L_1/L_2$) of the taper (Fig. 1b), which introduces the phase modulations and breaks the spatial parity-symmetry in the metasurface [5]. The forward transmission power spectra remain high, but broadband reflection spectra are highly sensitive to $L_1/L_2$.

Such contrast can be significantly enhanced with a second flip-faced triangular slot of asymmetric metasurface (Fig. 1c). Geometry-enabled asymmetric coupling coefficients to the transverse traveling guided mode show transmission contrast up to 25dB. The simulated phase profile shows that the bi-layer structure is crucial for eliminating the direct coupling between fundamental modes and maximizing the transmission contrast in the reciprocal system (Fig. S2). Adding the second layer transforms the sinusoidal phase profile into a square-wave (binary) phase profile, which is a crucial mode conversion mechanism from the zero-order mode to a higher-order mode in the forward direction[42–44].

In a single-layer asymmetric metasurface, the phase profile of the transmitted wave follows the sinusoidal distribution (varying between ± 0.4 π). The diffraction efficiency (DE) is defined as: $DE = J_m^2\left(\frac{\Delta\phi}{2}\right)$, where $J_m$ is the Bessel function, $m$ is the order of the mode, $\Delta\phi \leq 0.8$ π. The *minimal* zero-order DE of 10% can be achieved at $\Delta\phi = 0.8$ π. Numerically, Fig. S2(a) shows the sinusoidal phase profile of the single-tapered structure and induces the zero-order mode in forward excitation, which leads to the direct coupling between the input and output fundamental modes.

In the bi-layer asymmetric metasurface, however, the phase profile is square-wave (Fig. S2b). DE of coupling between input/output zero-order modes under square-wave phase profile is:

$$DE = |C_0|^2 = \frac{1}{2}(1 + \cos(\Delta\phi)) \tag{1}$$

Where the Fourier coefficient $C_m$ is defined as follows (m is the order of the mode):

$$C_m = \begin{cases} \frac{1}{2}(e^{i\Delta\phi} + 1), & m = 0 \\ \frac{1}{2}(e^{i\Delta\phi} - 1)\text{sinc}\left(\frac{m}{2}\right), & m \neq 0 \end{cases} \tag{2}$$

Therefore, DE can be zero when $\Delta\phi = \pi$ in the bi-layer structure. This means that the entire mode is converted from zero- to first-order mode. Fig. S2(b) shows the flipped bi-layer structure and resulting square-wave phase profile. In addition, when $\Delta\phi$ between unit cells is nearly $\pi$, the zero-order mode is eliminated, leaving only the first-order mode. Therefore, no backward transmission occurs due to the absence of the zero-order mode in both transmissions.

The proposed diffractive structure leverages a double-tapered bi-layer geometry to optimize phase distribution, ensuring the highest transmission contrast through perfect mode conversion. This approach effectively minimizes backward transmission by eliminating the zero-order mode in forward excitation, thereby achieving asymmetric transmission. In addition, asymmetric transmission for the double-tapered structure is maximized through parametric studies, as shown in Fig. S3. Among the design parameters, the gap between the larger and smaller taper ($G$), the smaller taper length ($L_1$), and the larger taper length ($L_2$) significantly affect the backward excited transmission and reflection spectra.

In addition, we optimized the insertion loss, phase shift, and asymmetric transmission through a few critical parameters in the unit cell of double-flipped taper slots, including the widths of the tapers ($W_{m,n}$), inter-taper gap ($G$), and the length of each taper ($L_{1,2}$). We fixed the lattice constant of 800 nm. Transmissions with forward ($T_F$) and backward ($T_B$) excitations are carefully mapped versus those design parameters (Fig. S1 and S3). The back reflection (or the transmission contrast $T_F/T_B$) is most sensitive to the distance between the double-flipped tapers. After fine-tuning the geometries, transmission contrasts upto 50dB can be achieved within the tens of nm range. The operation wavelength range can be effectively controlled by varying the parameters $G$. We verified that such transmission contrast is insensitive to the $L_{rect}$, so we can vary $L_{rect}$ to introduce phase gradience for the output wavefront. The simulation shows that optimal 55 dB transmission contrast can be achieved using the parameters $W_m$: 0.3 μm, $W_n$: 0.15 μm, $G$: 0.46 μm near the wavelength of 1.48 μm with 2D simulation.

Achieving asymmetric or directional light transport is not directly associated with broken reciprocity in multimode channels[45], but requires careful geometric engineering to maximize the contrast. We compare the mode conversions in single and bilayer metasurface in Fig. 1d, illustrating the critical role of the double-layer structure. Such suppression is not limited to the back excited normally incident waves but also results in broadband 50% back scattering reflection from circular scatterers with random sizes and positions (Fig. 1f). Compared to the control structure without the metasurface, the broadband reflection from the random scatterers was suppressed from 50% to less than 10% across the 300 nm wavelength range in near-infrared (Fig. 1e-f). Those scatterers randomly deflect light in different directions. We did a separate set of numerical simulations to obtain the incident angle-dependent forward and backward transmission (Fig. S4). The backward excited transmission reduces with incident angle. Within ±10° of incidence, backward excited transmission is kept below -10dB while forward transmission remains high (Fig. S4e).

To establish a theoretical foundation for asymmetric transmission, consider a two-port system where each port supports multiple diffraction orders due to the periodicity of the metasurface. Acting as a diffraction grating, the metasurface presented in the previous sections couples light into the zero-order mode ($m = 0$), referred to as Mode A, and the first-order diffraction modes ($m = \pm 1$), collectively represented as Mode B. The scattering matrix of such a system can be expressed as[3]:

$$\bar{\bar{S}} = \begin{bmatrix} S_{11}^{AA} & S_{11}^{AB} & S_{12}^{AA} & S_{12}^{AB} \\ S_{11}^{BA} & S_{11}^{BB} & S_{12}^{BA} & S_{12}^{BB} \\ S_{21}^{AA} & S_{21}^{AB} & S_{22}^{AA} & S_{22}^{AB} \\ S_{21}^{BA} & S_{21}^{BB} & S_{22}^{BA} & S_{22}^{BB} \end{bmatrix} \quad (3)$$

where the subscripts denote the port number, and the superscripts denote the mode number. Reciprocity requires that $\bar{\bar{S}} = \bar{\bar{S}}^T$, and energy conservation requires that $\bar{\bar{S}}^T \bar{\bar{S}}^* = \bar{\bar{U}}$, where $\bar{\bar{U}}$ is the unitary matrix.

If the system is excited from Port-1 with Mode-A, the output power measured *only* at Port-2 becomes $\left|S_{12}^{AA}\right|^2 + \left|S_{12}^{AB}\right|^2$. On the other hand, if the system is excited in the opposite direction from Port-2 with the same Mode-A, the output power measured *only* at Port-1 becomes $\left|S_{21}^{AA}\right|^2 + \left|S_{21}^{AB}\right|^2$. Reciprocity requires that $S_{12}^{AA} = S_{21}^{AA}$, however, in the absence of mirror symmetry along the longitudinal direction, $S_{12}^{AB}$ maybe different than $S_{21}^{AB}$. As a result, one can maximize the "asymmetric response" of a reciprocal system by setting $S_{12}^{AA} = S_{21}^{AA} = 0$, which, together with the reciprocity condition, gives the following scattering matrix:

$$\bar{\bar{S}} = \begin{bmatrix} S_{11}^{AA} & S_{11}^{AB} & 0 & S_{12}^{AB} \\ S_{11}^{BA} & S_{11}^{BB} & S_{12}^{BA} & S_{12}^{BB} \\ 0 & S_{21}^{AB} & S_{22}^{AA} & S_{22}^{AB} \\ S_{21}^{BA} & S_{21}^{BB} & S_{22}^{BA} & S_{22}^{BB} \end{bmatrix} \quad (4)$$

The 'efficiency' of this system can be further improved by setting $S_{11}^{AA} = S_{11}^{AB} = S_{12}^{BA} = S_{12}^{BB} = S_{22}^{AB} = S_{22}^{BB} = 0$, which would result in the following scattering matrix for a reciprocal and lossless system with an ideal "asymmetric response:"

$$\bar{\bar{S}} = \begin{bmatrix} 0 & 0 & 0 & 1 \\ 0 & 1 & 0 & 0 \\ 0 & 0 & 1 & 0 \\ 1 & 0 & 0 & 0 \end{bmatrix} \quad (5)$$

In this ideal scenario, if Mode-A is injected into Port-1, it will be completely converted into Mode B at Port-2. On the other hand, if Mode-A is injected into Port-2 it will be completely reflected as Mode-A at Port-2.

Within the above context, to further analyze the asymmetric response achieved by our metasurface design presented in Fig. 1, we consider excitation from Port 1, where the fundamental mode (Mode A) is converted to the first-order mode (Mode B) and transmitted efficiently to Port 2 (left panel of Fig. 2a). Conversely when the fundamental mode (Mode A) is excited from Port 2, it is predominantly reflected into the same mode at Port 2, resulting in strong backward reflection (right panel of Fig. 2a). This behavior arises from the asymmetric geometry of the metasurface, which enables directional mode coupling without violating reciprocity (Fig. 2b). The near-field distributions within a unit cell for Mode A excitation from Ports 1 and 2 are compared in the top panels of Fig. 2a, highlighting the distinct field profiles responsible for the asymmetric response. To further quantify this behavior, we performed far-field simulations for both Mode A and Mode B excitations from Port 1 (Fig. 2c) and Port 2 (Fig. 2d). These simulations allowed us to extract the elements of the scattering matrix, which describe the transmission and reflection characteristics of the system and demonstrate the reciprocity-constrained asymmetric response of the metasurface.

With the optimized unit cell geometry, the metasurface was integrated into a multi-mode waveguide defined on 250 nm silicon-on-insulator (SOI) substrate (Fig. 3a). Patterns with a set of geometric offsets (in the step of 10 nm, considering the minimum resolution of 7 nm of the E-beam lithography) are included in the layout to compensate for fabrication variations[46]. The right inset of Fig. 3a illustrates the details of the fabricated triangle metasurface with fine structures, with critical dimensions around 50nm. With the fundamental mode excited in the 10µm wide waveguide, full field simulation captures the in-plane field distribution with excitations from forward (port 1) and backward (port 2) excitations (Fig. 3b-c). Fig. 3b compares the cross-sectional profile across the waveguide, for the input (black), forward transmitted (orange) and back-excited transmitted (purple) modes. Fig. 3c indicates that, under forward excitation, the diffracted beam undergoes total internal reflection (TIR) on the boundary of multimode waveguide. In contrast, under backward excitation, the beam is reflected at the metasurface, effectively suppressing backward transmission. The transmitted and reflected power spectra are given in Fig. 3d-e (with monitors marked in Fig. 3c). The reflected spectra are experimentally verified, as the fundamental mode propagation is not susceptible to taper or coupler designs (detailed in Fig. 5).

We implemented another set of devices on a lower refractive index contrast silicon nitride platform (Fig. 4). Fig. 4a illustrates the test bed with optical microscope image of device under test (DUT1). The zoom-in image of the metasurface in multi-mode waveguide is given in Fig. 4b. We verified the design with full-field simulation of the mode profile in the waveguide, with independent excitations for forward (orange) and backward excitations (purple) (Fig. 4b-c). Opposite to the transmission results, the reflection monitor confirms that the reflected mode remains the fundamental mode with backward excitations (Fig. 4c), and forward reflection is quite minimal. Experimentally, we observed clear and consistent results for asymmetric reflection (upto 8dB, with 20 nm geometric offsets). Four sets of devices with identical design but different geometric offsets (-40, -20, 0, and 20 nm) were compared (Fig. 4e). To visualize such asymmetric reflection, we fabricated another set of devices with destinated reflection ports connected to vertical couplers (Fig. 4f). With forward excitation from port 1 on the left, no reflection was observed under top imaging infrared camera (Fig. 4g), while the back reflection leads to noticeable reflections (Fig. 4h), even with additional loss through the couplers. The loss of each coupler is carefully evaluated and compared across devices and verified to be ~ 8dB.

We also performed numerical simulation of the design's sensitivity to fabrication variation. Five different offsets were set upto 40 nm were given in Fig. S5a. These offsets have a limited impact on the forward transmission coefficient, with a range of -0.2 dB to -3 dB in the wavelength range of 1300-1600 nm (Fig. S5b). However, offsets reduce the highest contrast for backward transmission. Given 40nm geometric offset, the transmission dip is still below -35dB, indicating more than 98% of backward excited light was reflected (Fig. S5c).

To extend the utility of the proposed structure, we studied the coupler and interface to the single-mode waveguides or waveguides with different dimensions (Fig. 5). Two types of couplers (step couplers and conventional taper structures) were considered for converting mode A or B to the output waveguide. Simultaneous high forward transmission and high backward reflection can be translated from the metasurface level to the integrated photonic device level with step couplers (Fig. 5a, c, d). The forward transmission (converted to mode B) experiences high propagation loss in the taper converter/coupler (Fig. 5f), while high reflection remains the same with alternative taper designs, (Fig. 5g). We analyzed the mode-dependent insertion loss in taper in Fig. 5h. The incident angle on the boundary of multi-mode waveguide stays beyond the critical angle at the first interface (Fig. 5b, f), however, the relative incident angle reduces as the wave propagations along the taper, eventually coupling to the lossy/leaky wave. The conventional taper design cannot deliver asymmetric power transportation to the integrated photonic device level, as the component loss of the taper is also highly asymmetric (Fig. 5i, j). An additional simulation of the taper design-dependent transmission is provided in Fig. S6.

## CONCLUSIONS AND DISCUSSIONS

With optimized topologies and phase matching, we implemented asymmetric transmission on the integrated photonic platform using a phase-varying asymmetric metasurface. Asymmetric power transmission, insertion loss, and reflection contrast are enhanced with flipped bilayers. The forward insertion loss and back reflection improved from 1.5dB and 70% for single layers to 0.96dB and 80% for bilayers. The performance is consistent over the 200nm wavelength range. Following reciprocity, the low loss forward transmission is achieved by converting the incident fundamental mode to the high order mode, while the high reflection applies to the back-excited fundamental mode. In addition, integrating metalens with the asymmetric coupler design allows the focusing of the diffracted beams. This may reduce the footprint of the device for large-scale circuit integrations [47–49].

We experimentally implemented the device concept on both silicon and silicon nitride platforms and demonstrated the effectiveness of our proposed concept by comparing the reflection contrast and insertion loss over a broad range of wavelengths. The strong backward reflection through port 2 remains the same mode as excitations, as thus conventional taper designs are sufficient to validate asymmetric reflection. However, conventional taper design introduces additional loss for the forward transmitted high order mode. To maintain the low insertion loss of the forward transmission, a step-coupler may be involved to modify the mode size.

Table 1 summarizes the on-chip asymmetric responses achieved with metasurface. Most metasurface works focus on slow-varying geometric parameters (e.g., metalenses) for spot size modification[50,51]. Tapering or tilting the symmetric metasurface enables mode conversions in multi-mode wavelengths[52–54]. The footprint and effective length of the device can be minimized with phase-varying asymmetric metasurface design.

No backward transmission degradation occurred in a rectangular slab structure with infinite taper length (Fig. S6). However, spherical waves propagated in the taper waveguide because the taper length was too short to propagate as a plane wave (Fig. S6b). As a result, the short taper leads to higher transmission with backward excited waves (Fig. S6c-e). Therefore, we anticipate that by incorporating the metalens with an appropriate focal length can balance the insertion loss and device footprint. In addition, it's difficult to sustain backward suppression in a step coupler with a smaller width of waveguide (<3 µm) (Fig. S7a). The waveguide achieves the optimal transmission contrast of over 25 dB when no step coupler or taper is applied (Fig. S7b). Nevertheless, compared to other metasurface base structures, the subwavelength asymmetry allows low loss, broadband, and compact mode conversion, bringing the unique function of unidirectional reflection in multi-mode waveguides. The demonstrated device design may be integrated with chip-scale amplifiers and lasers, and thus find applications in photonic systems across a broad range of power requirements, from distributed gain LiDAR, complex photonic computing architecture, few-photon quantum communication to high-power RF-photonic remote sensing.

**Methods**

*Device characterizations:* A tunable laser source in the telecommunication C-band sends the TE polarized light to the on-chip grating coupler through a polarization controller and single-mode fiber. The output power was measured using the NEWPORT InGaAs photodiode (818-IG-L-FC/DB) and 1830-R optical power meter. For reflection measurement (Fig. 3 and 4), an additional circulator was added between the polarization controller and the device. The additional port of the circulator was connected to the optical power meter. More details of the device fabrication and characterizations of the asymmetric power transmission are provided in our previous work [35]. The fabrication offset sensitive Y junction insertion losses (methods) were characterized (Fig. S8) and subtracted from the total transmission to obtain the asymmetric metasurface-only response.

*Numerical simulation*: The optical field profile, transmission, and excitation spectra were performed by the 3D finite-difference-time-domain method (FDTD). The simulation was conducted in three steps: unit cell structure, metalens structure, and metalens integrated taper waveguide structure. For unit cell simulation, a silicon dioxide-based double-flipped taper slot structure was designed on the 250 nm thick Si layer in oxide. A pair of periodic boundary conditions were set in the x-axis direction to consider the metalens structure. A plane wave was used, and the direction of the electric field was set to be orthogonal to the periodic boundary condition direction. Metalens was designed based on unit cell simulation. A Gaussian mode source was used, considering the SMF excitation, and the boundary conditions were set to a perfect matching layer (PML) to account for a free-space radiation environment. Finally, a metalen has been integrated with waveguide simulation. Considering the sizes of the metalens and the waveguide, we employed an 11 μm to 0.5 μm taper structure, and the length of the taper was determined as 20 μm, considering the focal length. The excitation source and boundary conditions were set as Gaussian mode and PML boundary, respectively, consistent with the previous configuration.

**Supporting Information**.

Details of optical simulation and characterization can be found in the supporting information.

**Acknowledgment**

The SiN wafer with waveguide and metasurface were fabricated in the UCSB Nanofabrication Facility with deep-UV lithography. The Si photonic devices were fabricated at the University of Delaware Nanofabrication Facility (UDNF). This work was supported by the Defense Advanced Research Projects Agency with grant no. N660012114034, the Office of Naval Research with grant no. N00014-22-1-2486, the Air Force Office of Scientific Research with grant no. FA9550-22-1-0204, and the National Science Foundation with grant no. 2338546.

**Figures**

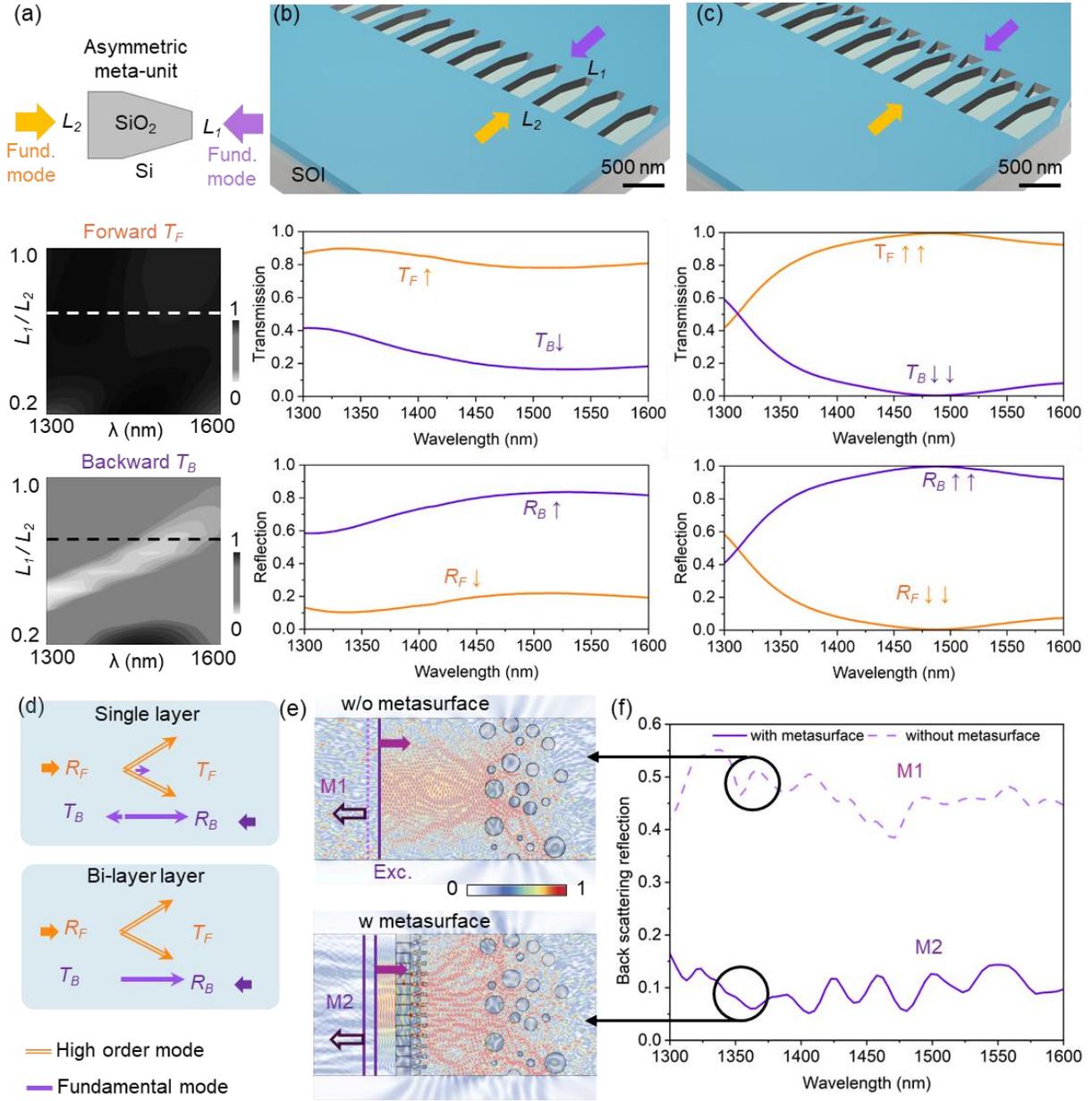

**Figure 1. Geometric optimization for high forward transmission and backward reflection and application scenario in multi-mode waveguide.** (a) The unit cell of the asymmetric metasurface defined on silicon-on-insulator (SOI) substrate, and corresponding forward (middle) and backward transmission (bottom) spectra with varying geometry asymmetry ($L_1$, $L_2$), where $L_1$ is the dimension on the shorter end and $L_2$ is the one for the wider end of the asymmetric meta-unit (lattice constant of 800 nm, silicon thickness of 250 nm). (b) Asymmetric metasurface exhibiting

higher forward transmission ($T_F$) and suppressed backward transmission ($T_B$). (c) Double flipped taper slot with broadband high contrast of transmission and reflection. $T_F$>90% and $R_B$>90% are achieved over 100 nm simultaneously. (d) Mode conversions with single-layer and bi-layer metasurface. (e) Exemplary application scene: suppression of back reflection. Intensity of electric field distribution of light scattering in a random media, triggering strong reflection (upto 50%), as recorded by the monitor (dashed line). Such reflection is greatly reduced after adding the asymmetric metasurface in (c), resulting in only 10% of reflection. (f) Backscattering reflection spectra with and without asymmetric metasurface.

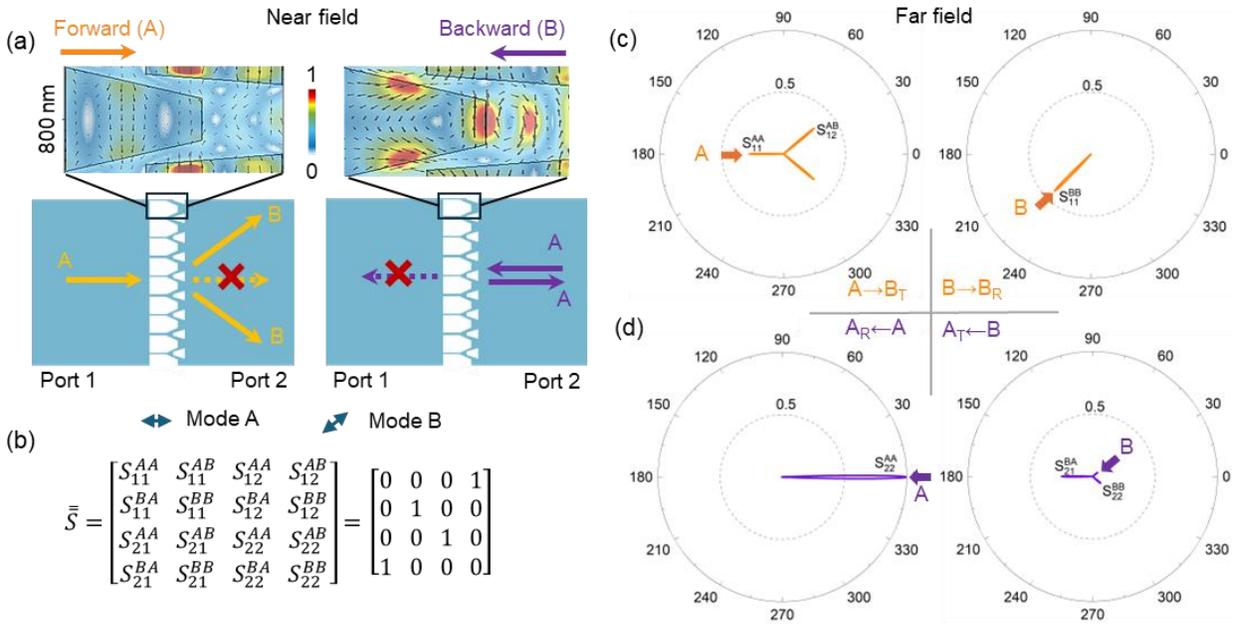

**Figure 2. Momentum and mode conversions.** (a) Mode momentum conversion analysis with forward (left) and back (right) excitations of guided modes towards the double-flipped asymmetric metasurface defined in the SOI substrates. Insets: electric field distribution in a unit cell with forward and backward excitation. (b) Correspondent transmission matrix for the dual modes (mode A and B are for zero and first-order diffraction given the lattice constant) and dual port system

(Port 1 and 2). Here the modes are defined through the propagation direction in the far field (detailed simulation provided by panels c and d). All the modes are in the same polarization. (c) Far-field response of the double-flip metasurface with forward excitation of mode A (left) and mode B (right), and (d) backward excitation of mode A (left) and B (right). High reflection to the same mode of excitations is identified for forward excitation with mode B and backward excitation with mode A. While forward excitation of mode A and backward excitation of mode B were converted to the other mode and resulted in high transmission.

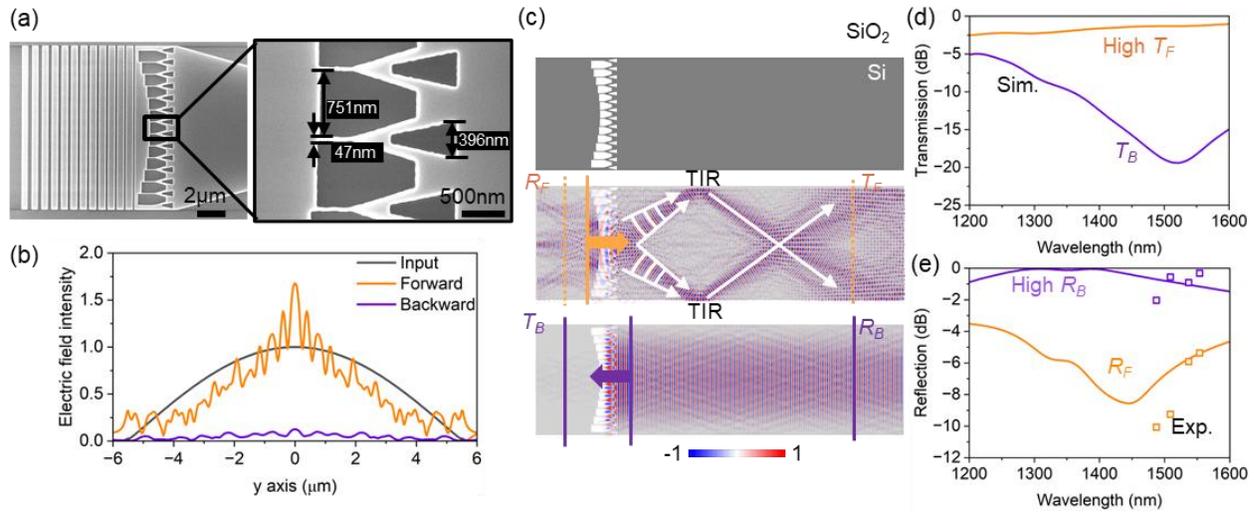

**Figure 3. Devices implementation**. (a) Scanning electron microscope image of the asymmetric metasurface-based metalens and grating coupler (scale bar: 2 μm). Right-inset zoom-in image of the detailed structure, with a critical dimension of 47 nm (scale bar: 500 nm). (b) Optical field intensity along the monitors across the metasurface (marked in c), with forward and backward excitations. (c) Top view of the asymmetric metasurface embedded multi-mode waveguide (top), optical mode profile (real part of the electric field) with forward (middle, light ray superimposed on the field distribution) and backward (bottom) excitation of the fundamental mode. TIR: total internal reflection. (d) Simulated and measured transmission (top) and (e) reflection with the excitations of the fundamental modes, compared the experimentally measured reflection (squares, setup detailed in Fig. 4a). Purple: forward. Yellow: backward excitations.

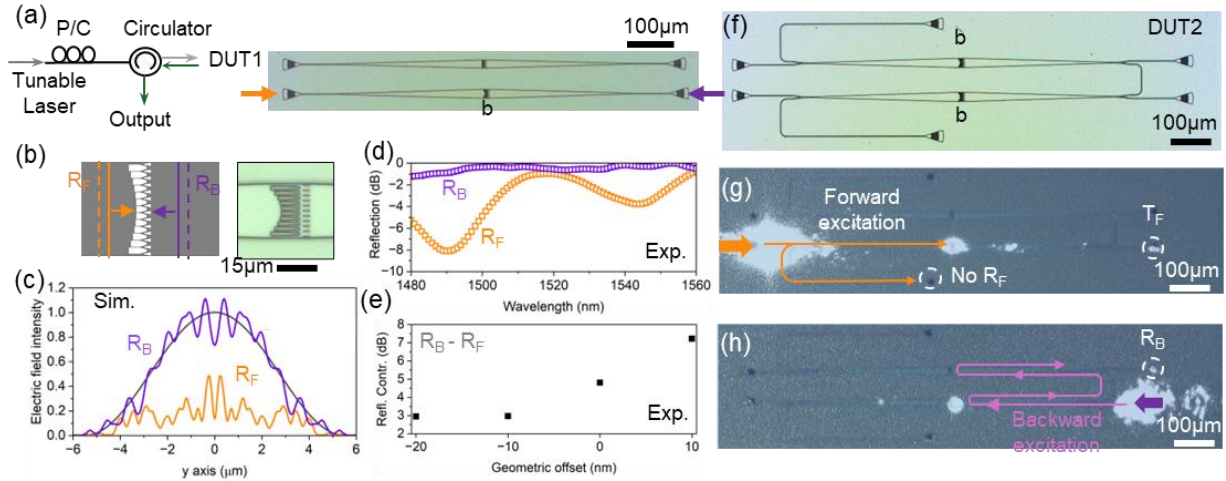

**Figure 4**. **Measurement of asymmetric reflection** in low refractive index contrast silicon nitride integrated photonic platform. (a) Apparatus for measuring the reflection from the device with two ports. P/C: polarization controller. The two-port devices (device under test DUT1) are two grating couplers connected to the double-flipped metasurface, with geometry optimized for wafer-scale manufacturing of silicon nitride. (b) Schematics and optical microscope image of the middle region of the metasurface. (c) Simulated reflection profile in the multi-mode waveguide without impacts from the taper. (d) Measured reflection with forward ($R_F$) and backward ($R_B$) excitations. (e) Measured reflection contrast with geometric offset (to compensate for fabrication variation). More than 10 nm fabrication offset is estimated. (f) DUT2 on the same chiplet, visualizing the asymmetric reflection. (g) Top view on the near-infrared camera with forward excitation of 1480 nm CW light. No output is identified on the reflection port. (h) The reflected light is identified with the backward excitation.

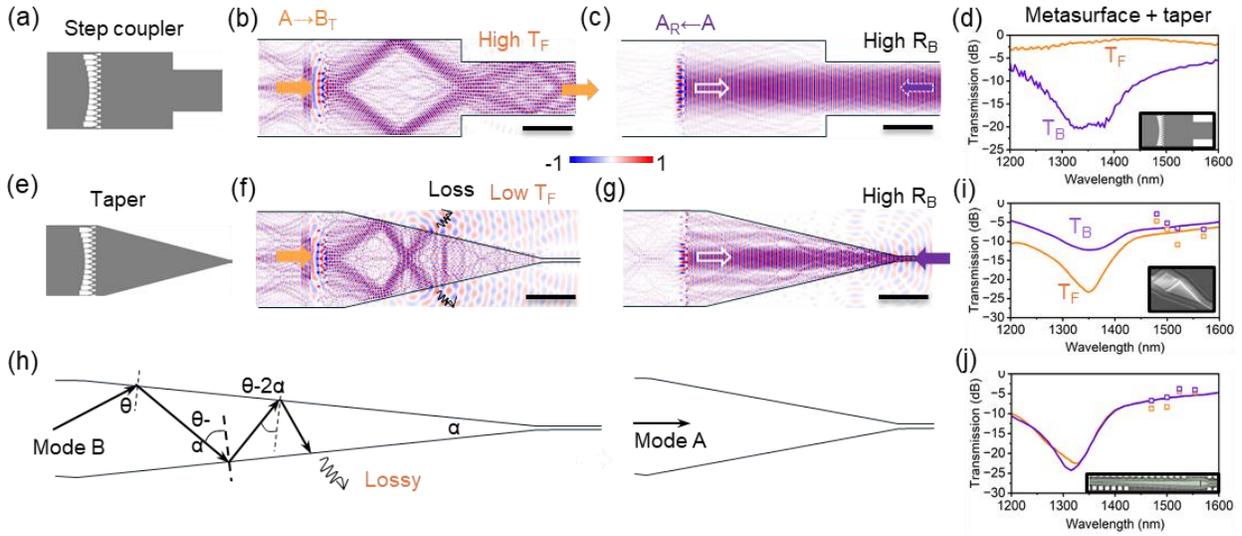

**Figure 5**. **Mode-dependent loss in mode-size converters and interface with single-mode waveguide.** (a) Step coupler-based mode-size converter, with embedded asymmetric metasurface. (b) Full-field simulation of the forward excitation in the multi-mode waveguide (10μm wide). The asymmetric metasurface converted the input fundamental mode to mode B. After total internal reflection on the interface, the light converged into the output waveguide with a smaller dimension. (c) Backward excitation towards the same structure results in the reflection of the same mode A. (d) High contrast asymmetric transmission with the step coupler. (e) Integration of the same metasurface with an exemplary taper. (f) The electric field distribution indicates the coupling to radiation, which results in high insertion loss and low coupling efficiency to the output waveguide. (g) High reflection remains the same as the step coupler. Scale bars: 5μm. (i) Inverted relation on the asymmetric transmission after integration with 20 μm long tapered coupler (curve: full field simulation. Dots: experimental results). (h) Ray-optic illustration of the mode-dependent insertion loss in the waveguide taper. (j) Experimental results for asymmetric metasurface with longer tapers and single mode waveguides inputs and outputs, exhibiting symmetric transmission.

**Table 1. Metasurface-based mode/mode size converters**

| Meta-unit | Length | Structure/Materials | Function | Reflection contrast | Loss | BW/$\lambda_0$ |
|---|---|---|---|---|---|---|
| **Asymmetric** | 2μm | 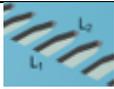 Dielectric | Fund. ↔ high order mode | -6.5 dB | 1.5dB | 300 nm/ 1550 nm |
| Double-flipped **asymmetric** | 2.5μm | 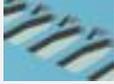 Dielectric | Fund. ↔ high order mode | -20 dB ~ -55 dB | 0.96dB | 200 nm/ 1550 nm |
| Gradient symmetric[32] | 20μm | 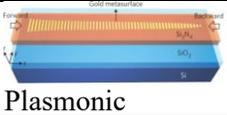 Plasmonic | Fund. ↔ high order mode | -20 dB | 3dB | 100 nm/ 2400 nm |
| Tilted symmetric[53] | 5μm | 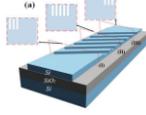 Dielectric | Fund. ↔ high order mode | None | <1dB | >20 nm /1550nm |
| Symmetric[54] | 15μm | 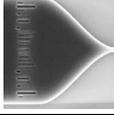 Dielectric | Fundamental mode size conversion | None | 1 dB | 200 nm/ 1550nm |
| Symmetric[50] | 30μm | 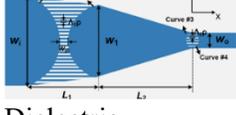 Dielectric | High order mode size conversion | None | <0.4dB | 200 nm/ 1550 nm |
| Symmetric[51] | 22μm | 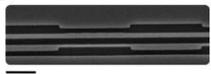 Dielectric | Fund. ↔ high order mode | None | 1.1dB | NA/ 1550nm |

Fund.: fundamental mode

**Supplementary file for**

**Broadband Low-loss Unidirectional Reflection On-chip with Asymmetric Dielectric Metasurface**

Heijun Jeong[1*†], Zeki Hayran[2†], Yuan Liu[3], Yahui Xiao[1], Hwaseob Lee[1], Zi Wang[1], Jonathan Klamkin[3], Francesco Monticone[2*] and Tingyi Gu[1*]

[1] Electrical and Computer Engineering, University of Delaware, Newark, DE 19716

[2] Electrical and Computer Engineering, Cornell University, Ithaca, NY 14853

[3] Department of Electrical and Computer Engineering, University of California, Santa Barbara, CA 93106, USA

*Email: heijun@udel.edu, francesco.monticone@cornell.edu, tingyigu@udel.edu

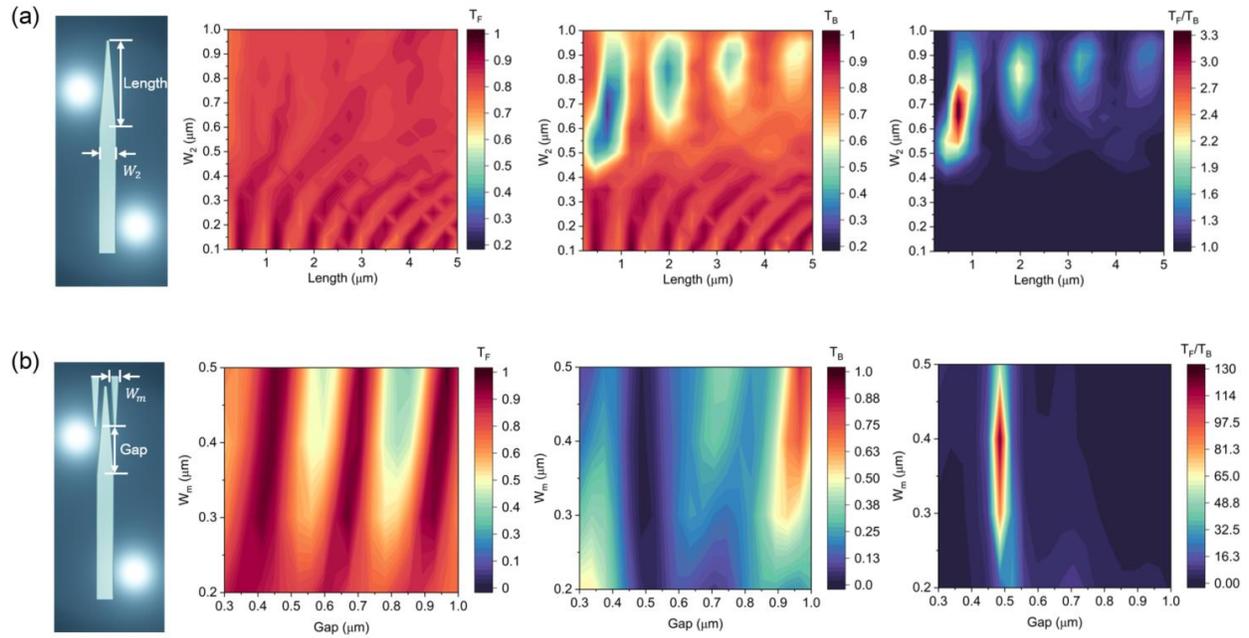

**Figure S1. Unit cell parameter optimization.** (a) Single taper structure, forward and backward transmission mapped verse $W_2$ and *Length*. (b) Proposed double flipped taper structure, forward and backward transmission mapped verse $W_m$ and $L_{rect}$.

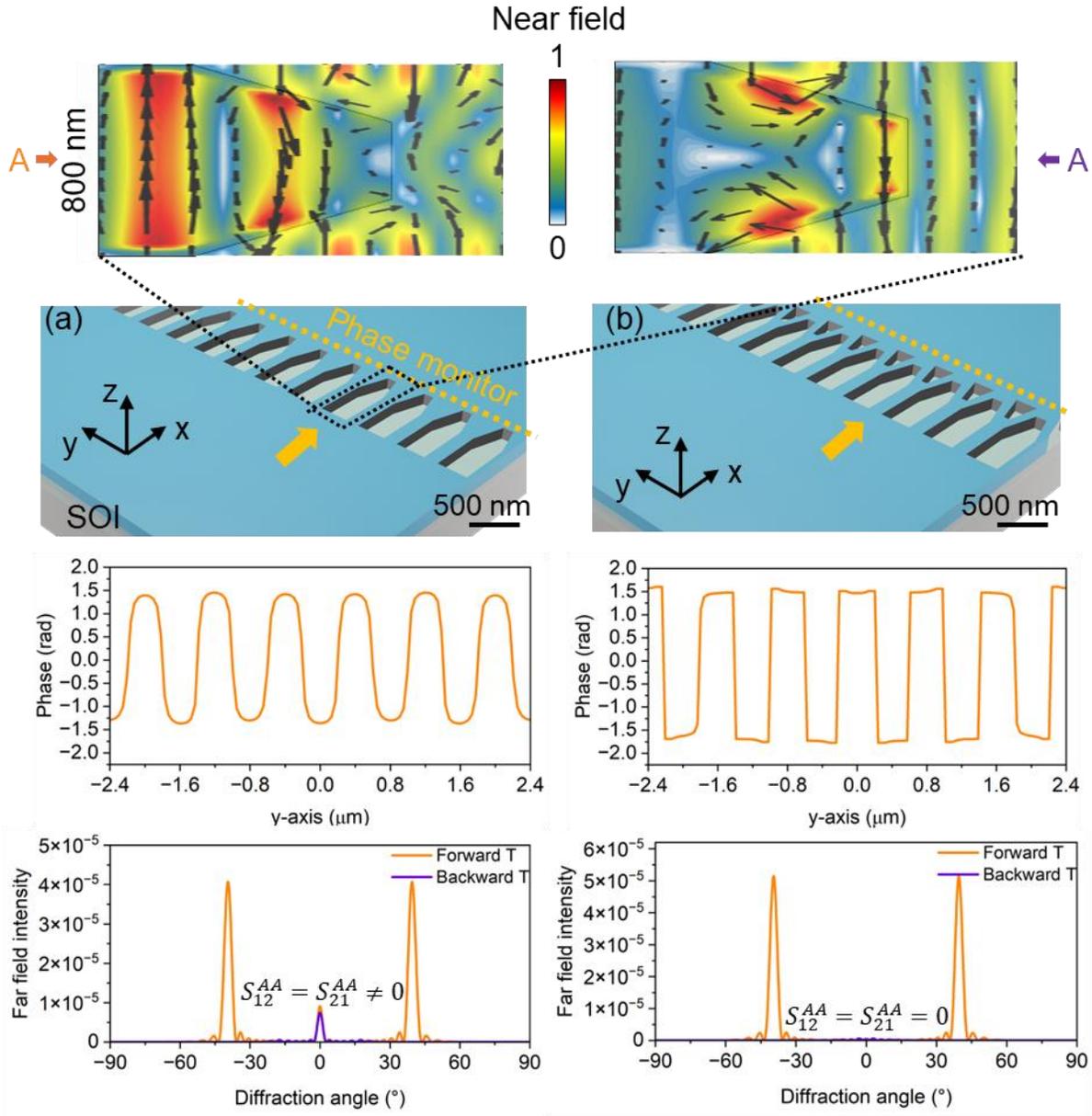

**Figure S2. Asymmetric transmission mechanism with diffraction mode conversion.** (a) Single tapered structure (top). Insets: electric field distribution in a unit cell with forward and backward excitation. Sinusoidal phase mapped forward transmission (middle) and zero-order mode exist in forward transmission, resulting in high backward transmission (bottom). The dashed orange line marks the position of the phase monitor, which is located beyond the metasurface for the forward transmission. (b) Double tapered structure (top). The square-wave phase was mapped for forward transmission (middle) and zero-order mode was eliminated in forward transmission, resulting in no backward transmission (bottom).

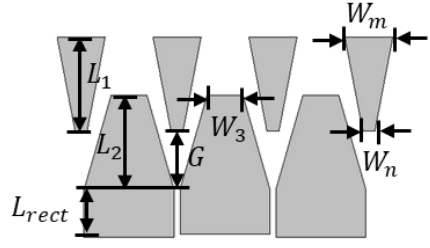
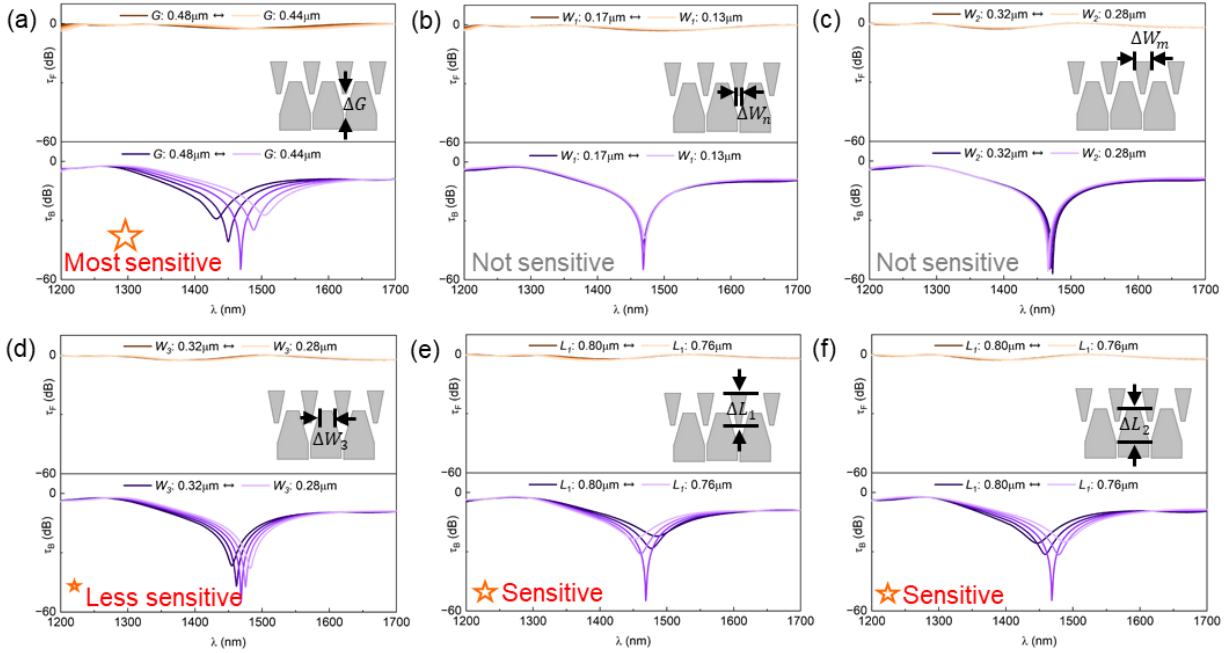

**Figure S3. Fine-tuning of unit cell geometry for achieving high contrast asymmetric coupling.** Forward transmission ($T_F$) and backward transmission ($T_B$) spectra at varying (a) gap distance ($G$), (b) $W_n$, (c) $W_m$, (d) $W_3$, (e) $L_1$ and (f) $L_2$. Top inset: detailed design parameters.

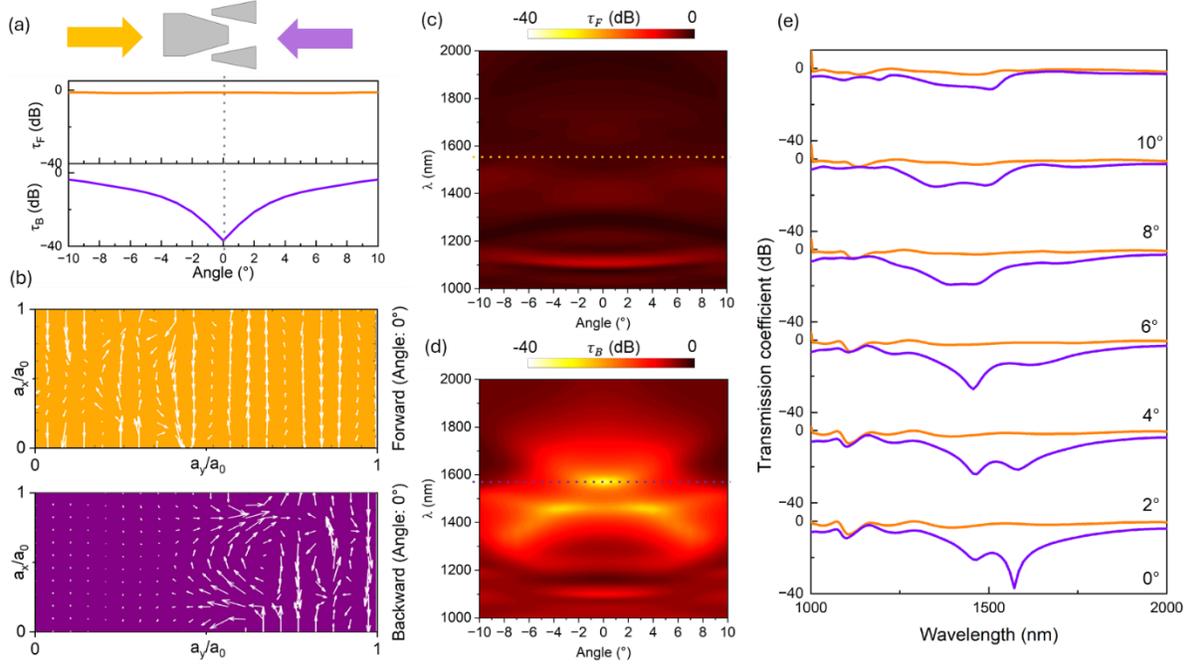

**Figure S4. Angle-sensitivity of the back-excited transmission.** (a) Forward and backward transmission spectra. (b) excitation schematics. (c) Angle-dependent forward and (d) backward transmission spectra. (e) Detailed transmission spectra within a unit cell. (d) Fiber-to-fiber transmission spectra of forward and back excitations. (e) transmission for forward and reverse excited waves with tilting angle to the normal incidences. Right-inset zoom-in result of the tilting angle of 10°.

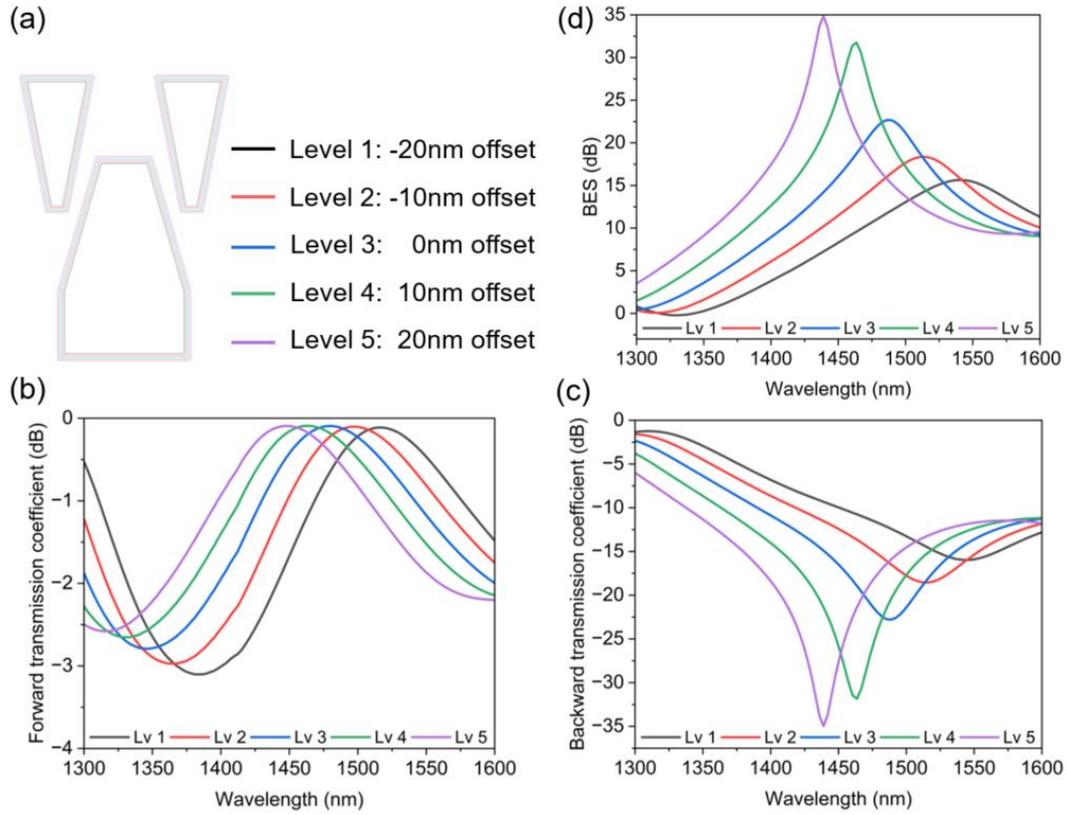

**Figure S5.** Unit cell simulation of geometric offsets. (a) Five different offsets with a 10 nm spacing, (b) forward transmission, (c) backward transmission, and (d) transmission contrast according to the offsets.

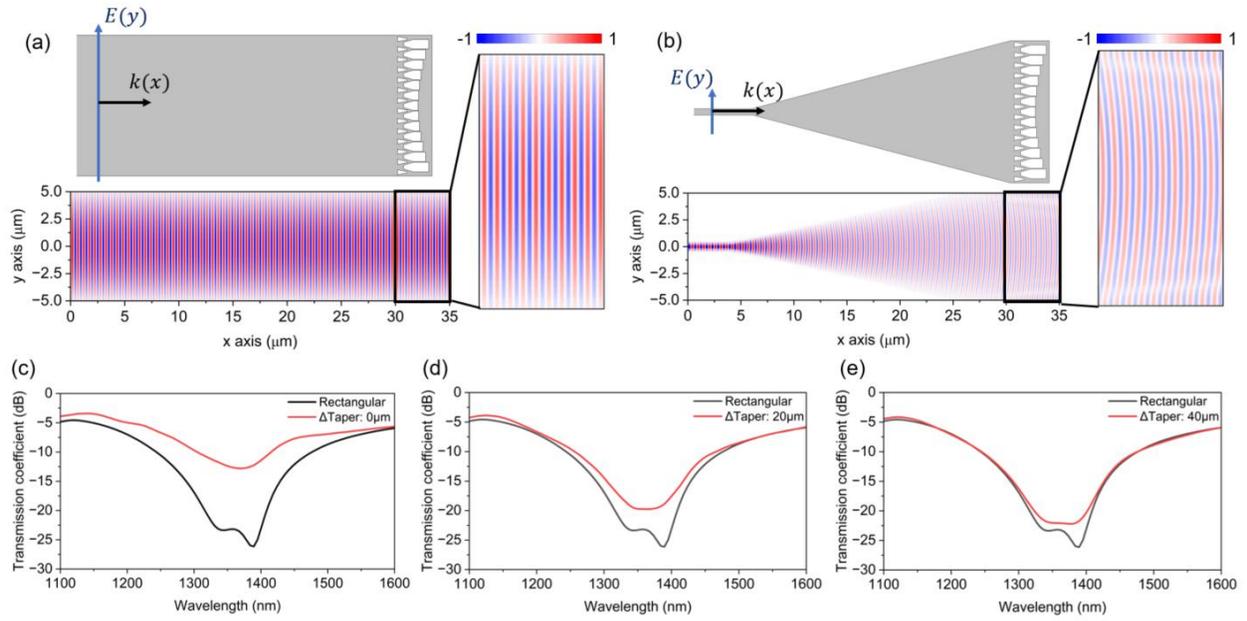

**Figure S6. Dependence of power transmittance contrast versus taper length.** (a) Full-field simulation of the electric field distribution ($E_y$) within a multi-mode waveguide (rectangular), and (b) from a single mode waveguide to multi-mode waveguide (taper). (c) Transmittance for backward excitation without taper (as a), (d) short taper (20 µm extended taper length as in b) and (e) long taper (40 µm extended taper length).

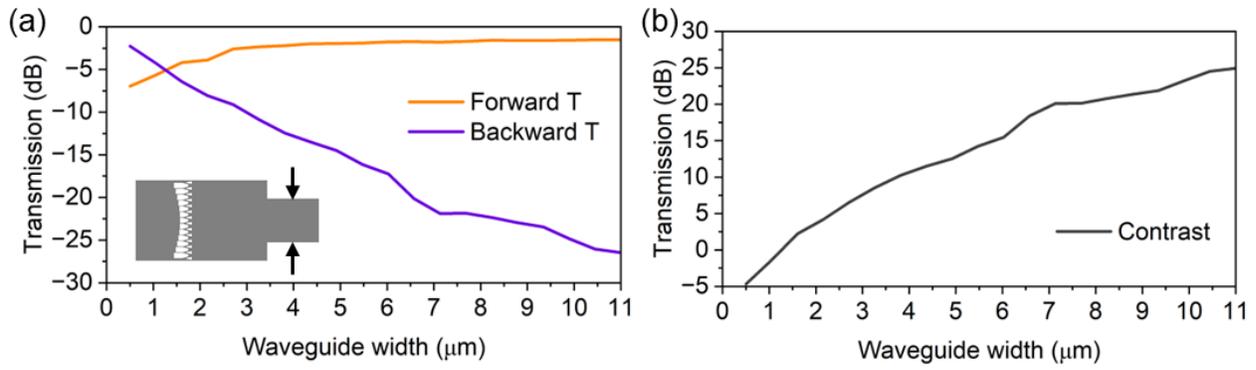

**Figure S7. Simulated transmission with varying step waveguide width**. (a) Forward and backward transmission. (b) Transmission contrast.

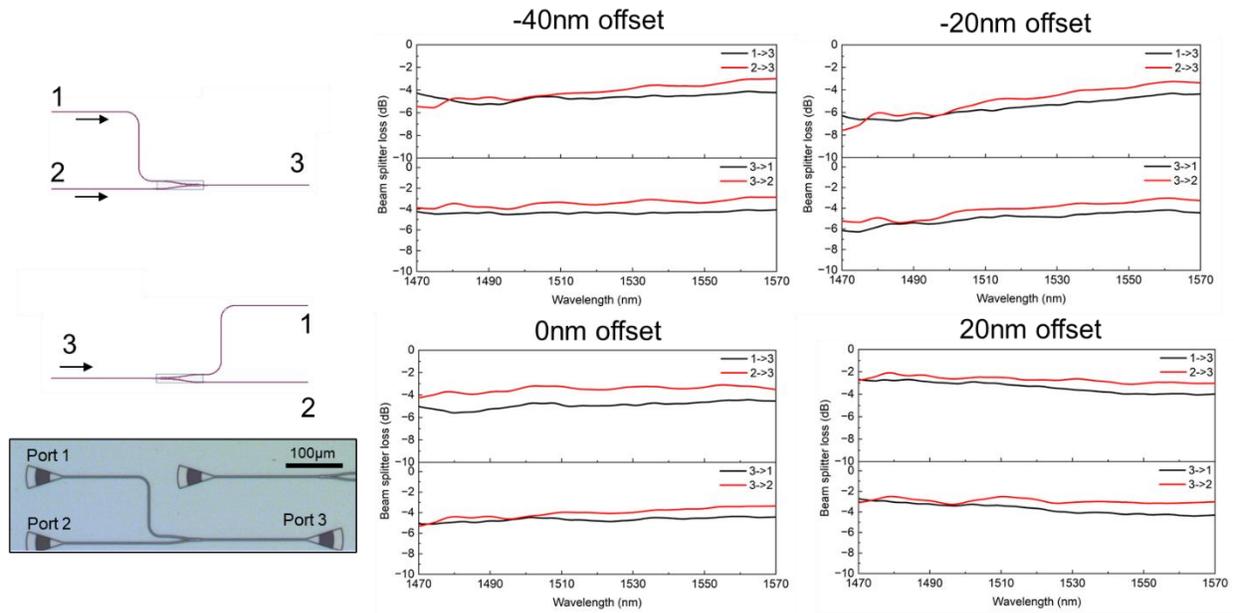

**Figure S8**. Y junction characterization versus geometric offsets on SiN wafer. Design with 20nm geometric offsets provide the lowest insertion loss, and thus the asymmetric metasurface design with 20nm offsets are selected for further investigation (results presented in Fig. 4). The error bars of measurement are estimated to be ~1dB based on the measurement.